\documentclass[manuscript]{acmart}

\AtBeginDocument{%
  \providecommand\BibTeX{{%
    \normalfont B\kern-0.5em{\scshape i\kern-0.25em b}\kern-0.8em\TeX}}}

\copyrightyear{2021} 
\acmYear{2021} 
\setcopyright{acmcopyright}\acmConference[RecSys '21]{Fifteenth ACM Conference on Recommender Systems}{September 27-October 1, 2021}{Amsterdam, Netherlands}
\acmBooktitle{Fifteenth ACM Conference on Recommender Systems (RecSys '21), September 27-October 1, 2021, Amsterdam, Netherlands}
\acmPrice{15.00}
\acmDOI{10.1145/3460231.3474238}
\acmISBN{978-1-4503-8458-2/21/09}

\begin{document}

\title{Next-item Recommendations in Short Sessions}

\author{Wenzhuo Song}
\authornote{This work was conducted during his visit at Macquarie University.}
\affiliation{%
  \institution{Jilin University}
  \streetaddress{2699 Qianjin Street}
  \city{Changchun}
  \country{China}}
\email{songwz17@mails.jlu.edu.cn}
\author{Shoujin Wang}
\authornote{Corresponding author.}
\affiliation{%
  \institution{Macquarie University}
  \streetaddress{Balaclava Rd, Macquarie Park NSW 2109}
  \city{Sydney}
  \country{Australia}}
\email{shoujin.wang@mq.edu.au}

\author{Yan Wang}
\authornotemark[2]
\affiliation{%
  \institution{Macquarie University}
  \streetaddress{Balaclava Rd, Macquarie Park NSW 2109}
  \city{Sydney}
  \country{Australia}}
\email{yan.wang@mq.edu.au}
\author{Shengsheng Wang}
\authornotemark[2]
\affiliation{%
  \institution{Jilin University}
  \streetaddress{2699 Qianjin Street}
  \city{Changchun}
  \country{China}}
\email{wss@jlu.edu.cn}

\renewcommand{\shortauthors}{Song, et al.}

\begin{abstract}
The changing preferences of users towards items trigger the emergence of \textit{session-based recommender systems (SBRSs)}, which aim to model the dynamic preferences of users for next-item recommendations. However, most of the existing studies on SBRSs are based on long sessions only for recommendations, ignoring short sessions, though short sessions, in fact, account for a large proportion in most of the real-world datasets. As a result, the applicability of existing SBRSs solutions is greatly reduced. In a short session, quite limited contextual information is available, making the next-item recommendation very challenging. To this end, in this paper, inspired by the success of few-shot learning (FSL) in effectively learning a model with limited instances, we formulate the next-item recommendation as an FSL problem. Accordingly, following the basic idea of a representative approach for FSL, i.e., meta-learning, we devise an effective SBRS called \textit{\textbf{IN}ter-\textbf{SE}ssion collaborative \textbf{R}ecommender ne\textbf{T}work (INSERT)}
for next-item recommendations in short sessions. With the carefully devised local module and global module, INSERT is able to learn an optimal preference representation of the current user in a given short session. In particular, in the global module, a similar session retrieval network (SSRN) is designed to find out the sessions similar to the current short session from the historical sessions of both the current user and other users, respectively. The obtained similar sessions are then utilized to complement and optimize the preference representation learned from the current short session by the local module for more accurate next-item recommendations in this short session. Extensive experiments conducted on two real-world datasets demonstrate the superiority of our proposed INSERT over the state-of-the-art SBRSs when making next-item recommendations in short sessions. 
\end{abstract}


\maketitle

\section{Introduction}
In the real world, a user's preference towards items usually changes over time, leading to the need to model the user's dynamic and more recent preference for more accurate recommendations. To this end, session-based recommender systems (SBRSs) have emerged in recent years to model users' dynamic and short-term preferences for next-item recommendations \cite{hidasi2015session,wang2019survey}. Specifically, given a user's session context, e.g., a few selected items in an online transaction or a shopping basket \cite{wang2020intention2basket}, an SBRS aims to predict the next item in the same session that the user may prefer.

Most of the existing studies \cite{liu2018stamp,wang2018attention} on SBRSs focus on long sessions only for next-item prediction, ignoring \textit{short sessions.} Following an existing work \cite{qiu2020exploiting}, \textit{short sessions} and \textit{long sessions} refer to the sessions containing no more than five items and those containing more than five items, respectively. The number of items contained in a session is referred to as the \textit{length} of the session. Specifically, the existing studies often follow a common practice to filter out short sessions during the data pre-processing to make the next-item prediction less challenging \cite{wang2020modelling,wang2019modeling}. This is because a short session contains only very few items, and thus the contextual information embedded in it is very limited, making the prediction highly challenging.

\begin{figure}
\centering
\includegraphics[width=0.9\columnwidth]{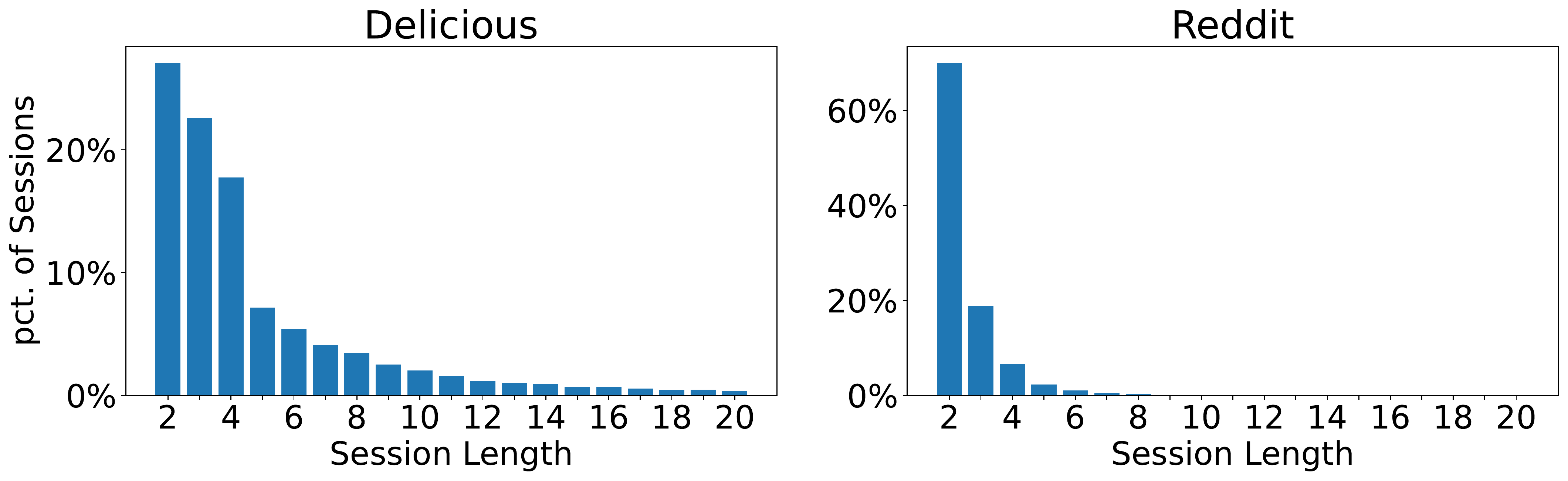}
\caption{Short sessions account for the majority in two real-world datasets ``Delicious'' and ``Reddit''.}
\Description{Two well-known real-world session-based datasets, ``Delicious'' and ``Reddit'', have 64\% and 97\% sessions with lengths less than 5, respectively.}
\label{fig_sesslen}
\end{figure}

However, ignoring short sessions greatly reduces the applicability of SBRSs in real-world cases. In practice, short sessions usually account for a large proportion of a dataset. For example, as depicted in Fig.~ \ref{fig_sesslen}, two well-known real-world datasets ``Delicious'' and ``Reddit'' have 64.03\% and 96.95\% short sessions, respectively. 

The above analysis and observation reveal that significant gaps w.r.t. making next-item recommendations in short sessions exist in the proposed solutions on SBRSs. According to the contextual information utilized for next-item recommendations, SBRSs can be roughly divided into \textit{single-session-based SBRSs} and  \textit{multi-session-based SBRSs.} Single-session-based SBRSs  \cite{hidasi2015session,wang2018attention} make recommendations based on the current session only, and hence \textbf{(Gap 1)}\textit{ the available information that can be used for recommendations is very limited due to the limited number of items in a short session.} This makes it very difficult to fully understand the user's preference, and thus it decreases the performance of recommendations. 

In order to leverage more contextual information to improve the recommendation performance, multi-session-based SBRSs take other sessions into account for the prediction of the next item in the current session. Specifically, some multi-session-based SBRSs \cite{quadrana2017personalizing,sun2019can} incorporate historical sessions of the user (called \textit{current user}), who generates the current session, to alleviate the problem of insufficient information in the current short session to some extent. However,\textbf{ (Gap 2)}\textit{ they ignore the rich information of other users similar to the current user}, which is intuitively helpful for those users who do not have enough historical sessions. 
Furthermore, other multi-session-based SBRSs \cite{ludewig2018evaluation} incorporate the sessions similar to the current session from other users into the next-item prediction in the current session. They first simply represent the current session with a fixed-length one-hot vector or the mean-value vector of embeddings of the items included in the session. Then, they use the representation of the current session as a key to retrieve a few sessions similar to the current session from the whole dataset as a reference for the prediction of the next item in the current session. However, they often obtain some irrelevant sessions due to the oversimplified session representation and the limited information in the key. Hence, \textbf{(Gap 3)} \textit{ it is still a problem that how to effectively find out those really relevant and useful historical sessions of the current session and then incorporate them for the next-item prediction in the current short session.}

In this paper, we bridge the above three gaps by proposing an \textit{\textbf{IN}ter-\textbf{SE}ssion collaborative \textbf{R}ecommender ne\textbf{T}work (INSERT)} to effectively find out those sessions similar to the current session to complement the limited information in it for next-item recommendations in short sessions. 
The design of INSERT is inspired by the success of few-shot learning (FSL) methods in learning a model for a task with a few instances  \cite{wang2020generalizing}. 
Specifically, following the basic idea of one representative FSL approach (i.e., meta-learning) \cite{peng2020comprehensive,yao2019hierarchically, yao2019automated}, INSERT effectively infers the user preference for recommending the next item in a short session by utilizing not only the information from the limited items in the current short session but also the learned useful prior knowledge from other similar sessions. 
Specifically, INSERT contains three modules: 
(1) a \textit{local module} to infer the user's preference from the current short session, (2) a \textit{global module} to learn useful prior knowledge from other sessions, including both the current user's and other users' historical sessions, and (3) a \textit{prediction module} to first modulate and optimize the inferred preference of the local module according to the prior knowledge learned by the global module and then predict the next item based on the optimized preference. In particular, in the global module, we design a similar session retrieval network (SSRN) to precisely retrieve those sessions similar to the current session from the current user and other users. 
The main contributions of this work are highlighted as follows:

\begin{itemize}
\item We propose an \textit{inter-session collaborative recommender network (INSERT)} for effective next-item recommendations in short sessions. To the best of our knowledge, we are the first to explicitly and specifically focus on next-item prediction in short sessions and the corresponding gaps. 


\item For the first time, we formulate the next-item recommendation in sessions as a few-shot learning (FSL) problem to target the particular gaps when performing next-item recommendations in short sessions. In particular, given a short session, we design both local and global modules to learn users' optimal preferences for accurate next-item recommendations effectively.

\item We conduct extensive experiments on two real-world transaction datasets, and the results show the superiority of our proposed method over the state-of-the-art approaches for next-item recommendations when the sessions are short.
\end{itemize}

\section{Related Work}

The existing studies on SBRSs have two groups: (1) single-session-based SBRSs, and (2) multi-session-based SBRSs.  

\textbf{Single-session-based SBRSs.} 
Single-session-based SBRSs make next-item recommendations based on the current session only. 
Recurrent neural networks (RNN) based approaches model the sequential dependencies among a sequence of items within the current session for the next-item prediction in it \cite{hidasi2015session,hidasi2018recurrent}. 
However, they are based on a rigid order assumption that any adjacent items within a session are highly sequentially dependent, which may not be true in most of the real-world cases \cite{wang2019sequential}. 
To relax this rigid order assumption, attention-based SBRSs build an attentive embedding for a given session context by assigning larger weights to those items in the current session that are more relevant to the next-item prediction \cite{li2017neural,liu2018stamp}. However, attention-based SBRSs easily bias a few popular items while ignoring others. 
More recently, graph neural networks (GNN) have been widely applied into SBRSs to learn the complex transitions between items within sessions for accurate next-item recommendations \cite{qiu2019rethinking,wu2019session,xu2019graph,ma2019memory}. However, they may not perform well in predicting next-item in short sessions because the complex transitions mainly exist in long sessions.  



\textbf{Multi-session-based SBRSs.} Multi-session-based SBRSs incorporate other sessions to complement the information in the current session for next-item recommendations in it. Some multi-session-based SBRSs, e.g., session-aware recommender systems \cite{wang2019survey,Latifietal21}, incorporate the historical sessions of the current user. For example, both hierarchical RNN (HRNN) \cite{quadrana2017personalizing} and inter- and intra-session
RNNs (II-RNN) \cite{ruocco2017inter} first employ a session-level RNN and an item-level RNN to encode a sequence of historical sessions of the current user and a sequence of items in the current session, respectively, and then combines the outputs from both RNNs to predict the next item in the current session. In another similar work \cite{sun2019can}, only those identified relevant historical sessions of the current user are incorporated for the next-item prediction in the current session.  
However, these methods ignore the useful information from other users who have similar preferences and shopping behaviours to the current user. 

Different from the above methods, other multi-session-based SBRSs incorporate the sessions similar to the current session of the current user from other users for next-item recommendations in the current session. K nearest neighbour based approaches \cite{ludewig2018evaluation,garg2019sequence} first find out those sessions in the dataset that are similar to the current session, and then take them as a reference when predicting the next item in the current session. However, due to the oversimplified similarity calculation method used in these approaches, it is hard to precisely find out the really similar sessions. Later, memory networks have been employed to retrieve the sessions similar to the current session of the current user from other users  \cite{wang2019collaborative,pan2020intent}. These approaches mainly focus on those similar sessions that happened recently, ignoring others.  
Recently, GNN has been utilized to map all sessions into a graph and then absorb the useful information from other sessions to help with the next-item prediction in the current session \cite{wang2019collaborative,pan2020intent}. However, complex transitions tend to exist in long sessions instead of short sessions so that the strength of GNN becomes plain in making recommendations in short sessions.

\section{Problem Statement}


Let $U=\{u_1,...,u_{n}\}$ denote the set of all $n$ users in the dataset, and $V=\{v_1,...,v_m\}$ denote the set of all $m$ items in the dataset. Each user $u\in U$ has a sequence of sessions $S_u=\{s_1^u,...,s_{|S_u|}^u\}$, e.g., $u$'s e-commerce transactions, where the subscript of each session denotes its order in $S_u$ w.r.t. its occurring time. 
The $l$-th session of $u$, i.e., $s_l^u\in S_u$, consists of a sequence of items, i.e., $s_l^u=\{v_1^{u,l},...,v_{|s^{u}_l|}^{u,l}\}$,  ordered by its occurring time, e.g., when it was clicked.







For a target item $v_t^{u_c,l}$ in the current session $s_l^{u_c}$ of the current user $u_c$ to be predicted, all the items occurring prior to $v_t^{u_c,l}$ in the current session $s_l^{u_c}$ form the intra-session context $C_{ia} =\{v_1^{u_c,l},...,v_{t-1}^{u_c,l}\}$ of $v_t^{u_c,l}$. Given $C_{ia}$, an SBRS aims to predict the next item $v_t^{u_c,l}$ in $s_{l}^{u_c}$. Specifically, a probabilistic classifier is trained to predict a conditional probability $p(v|\{C_{ia},{S}\})$ over each of the candidate item $v\in V$, where ${S}=\{S_u|u\in U\}$ contains all the historical sessions in the training set. Finally, those items with the top-K conditional probabilities are selected to form the recommendation list.


\section{Inter-session Collaborative Recommender Network}
\begin{figure*}[ht]
\centering
\includegraphics[width=1\textwidth]{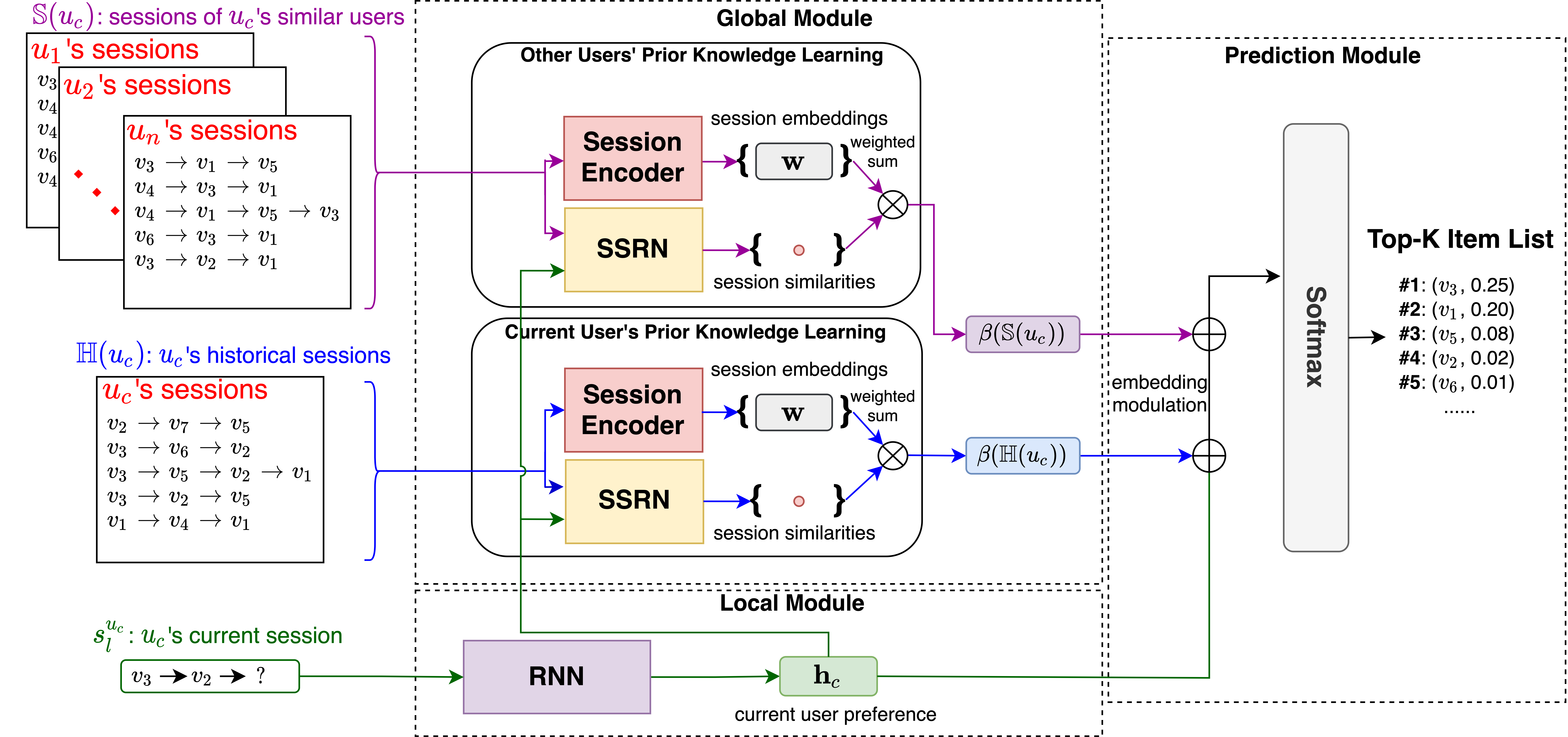}
\caption{The architecture of INSERT.}
\Description{INSERT contains a local module based on GRU to model the preference of the user in the current session, a global model based on SSRN and Session Encoder to learn useful prior knowledge from other historical sessions, and a prediction module based on MLP to modulate and optimize the inferred preference of the local module according to the knowledge learned by the global module and output a top-k item list. }
\label{fig_framework}
\end{figure*}

In this section, we introduce the proposed inter-session collaborative recommender network (INSERT). 
INSERT is inspired by the recent advances of few-shot learning, which aim to generalize to a new task containing only a few labelled examples \cite{wang2020generalizing}. In this work, we formulate next-item prediction in short sessions as a few-shot learning (FSL) problem by regarding a task as estimating the user's dynamic preference in the session. To overcome the challenge of unreliable user preference inference caused by limited items in $C_{ia}$, we employ the idea of one of the representative FSL approach, i.e., meta-learning, to effectively infer the user preference for recommending the next item in a short session \cite{peng2020comprehensive}. 
The key idea of the proposed INSERT framework is that 
by utilizing the learned useful prior knowledge from other sessions, we can effectively constrain the hypothesis space of the representations of the user preference given limited items in the current short session \cite{wang2020generalizing}. 

As depicted in Figure \ref{fig_framework}, INSERT contains (1) a \textit{local module} to infer the representations of the user preference based on $C_{ia}$ in the current session, (2) a \textit{global module} to learn useful prior knowledge from other sessions, including both the current user's and other users' historical sessions, and (3) a \textit{prediction module} to modulate and optimize the inferred preference of the local module according to the prior knowledge learned by the global module and then predict the next item based on the optimized preference. 
Specifically, the local module first learns a representation of the user preference, i.e., $\mathbf{h}_c$, based on $C_{ia}$ in the current short session. 
The user preference $\mathbf{h}_c$ can be used for next-item recommendation in the short session by applying a Multilayer Perceptron and a softmax layer:
\begin{equation}
    p(v|C_{ia})=softmax(MLP(\mathbf{h_c})),
    \label{eq:pred}
\end{equation}
where $v\in V$ denotes a candidate next item in the current session to predict.
However, $\mathbf{h}_c$ is not a reliable inference of the user preference due to the limited supervised information in $C_{ia}$. 
To alleviate this problem, the global module learns prior knowledge by retrieving useful information from other sessions, and the prediction module modulates and optimizes $\mathbf{h}_c$ based on a feature-wise modulation function $\psi$:
\begin{equation}
    \psi(\mathbf{h}_c,S)=\mathbf{h}_c+\beta(S),
    \label{modulation}
\end{equation}
where $\beta(S)$ is a $d$-dimensional vector representation of the prior knowledge learned based on $S=\{S_u|u\in U\}$, which is the set of all sessions in the training set. Thus, Equation \eqref{eq:pred} is rewritten as:
\begin{equation}
    p(v|C_{ia}, S)=softmax(MLP(\psi(\mathbf{h}_c,S))).
    \label{eq:pred2}
\end{equation}


\subsection{Local Module} 
The intra-session context $C_{ia}$ from the current session reflects the current preference of the current user $u_c$, which is the crucial information for next-item prediction. Specifically, we first embed each item $v_i$ in $C_{ia}$ to a $d$-dimensional vector representation $\mathbf{x}_i$, and then feed $\mathbf{x}_i$ into an RNN built on gated recurrent units (GRU):
\begin{equation} \label{eq1}
\begin{split}
\mathbf{h}_{i}=&GRU(\mathbf{x}_{i},\mathbf{h}_{i-1}),
\end{split}
\end{equation}
where $1\leq i\leq |C_{ia}|$. The first hidden state $\mathbf{h}_0$ is initialized with a zero vector. 
For each item in $C_{ia}$, we regard the output $\mathbf{h}_i$ of the corresponding GRU as the embedding of user preference at $i$ because the GRUs can automatically extract useful features from $v_i$ and the items previous to $v_i$ in this session. Besides, as a recursive model based on RNN, the local module can generate user preference embeddings when new items come and preserve the sequential patterns in the sessions. In this work, 
we use the most recent user preference embedding at $|C_{ia}|$ as the user's current preference used for next-item prediction in $s_l^{u_c}$, i.e., 
\begin{equation}
    \mathbf{h}_{c}=\mathbf{h}_{|C_{ia}|}.
\end{equation}


\subsection{Global Module}

To effectively learn useful prior knowledge from other sessions, we first form two candidate similar session sets as the inputs of the global module. The first is the set of $u_c$'s previous sessions $\mathbb{H}(u_c)$ and the second set $\mathbb{S}(u_c)$ contains the sessions of a few other users who have similar preferences with $u_c$. This design assumes that users with similar preferences are more likely to have sessions similar to the current session. 
The two candidate similar session sets are used as the inputs of two modules in the global module, i.e., Current User's Prior Knowledge Learning Module (CUPKL) which takes $\mathbb{H}(u_c)$ as input and Other Users' Prior Knowledge Learning Module (OUPKL) which takes $\mathbb{S}(u_c)$ as input. 
Given the user preference $\mathbf{h}_c$ in the current session, the CUPKL and OUPKL aim to learn prior knowledge from $\mathbb{H}(u_c)$ and $\mathbb{S}(u_c)$, respectively. They have the same architecture but different outputs, i.e., prior knowledge $\beta(\mathbb{H}(u_c))$ in $u_c$'s historical sessions and $\beta(\mathbb{S}(u_c))$ in the sessions of other users, which are used to modulate the user preference learned in the local module. 
In particular, 
both the CUPKL and OUPKL modules contain a similar session retrieval network (SSRN) to calculate session similarities and a Session Encoder to encode user preferences in candidate similar sessions. 
In the following, we will introduce the global module in detail. 

\subsubsection{Forming Candidate Similar Session Sets}
Retrieving similar sessions from the entire dataset will result in irrelevant sessions and high computational burden. To make it more precisely and efficiently, we use two similar candidate sets for the model to retrieve: one is the set of the sessions of $u_c$'s sessions prior to $s_l^{u_c}$:
\begin{equation}
    \mathbb{H}(u_c) = \{s_1^{u_c},...,s_{l-1}^{u_c}\},
\end{equation}
and the second is a set of the sessions of a few users who have similar preferences to the current user. Specifically, we first select $u_c$'s most similar users as those who interacted with most of the same items that $u_c$ has interacted with. Mathematically, we calculate the similarity between each user $u_\tau$ and $u_c$ by:
\begin{equation}
    sim_u(u_\tau,u_c)=\frac{|\Omega_\tau \cap \Omega_c|}{|\Omega_\tau|\times |\Omega_c|}
    \label{eq:usersim},
\end{equation}
where $u_\tau\in U$, $\tau\neq c$, and $\Omega_\tau$ is the set of items interacted by $u_\tau$. Then, we form a set of candidate similar sessions $\mathbb{S}(u_c)$ with the sessions of $u_c$'s $N$ most similar users in the training set\footnote{$N$ is empirically set to 10 in this paper since we found its performance with a larger $N$ did not change significantly but increase the running time.}. 
Accordingly, Equation \eqref{modulation} is rewritten as 
\begin{equation}
    \psi(\mathbf{h}_c,\mathbb{H}(u_c), \mathbb{S}(u_c))=\mathbf{h}_c+\beta(\mathbb{H}(u_c))+\beta(\mathbb{S}(u_c))
    \label{modulation2}
\end{equation}
where $\mathbb{H}(u_c)$ and $\mathbb{S}(u_c)$ represent the session sets of $u_c$ and his similar users, respectively.


\subsubsection{Similar Sessions Retrieval Network (SSRN)} 

\begin{figure}[t]
\centering
\includegraphics[width=0.95\textwidth]{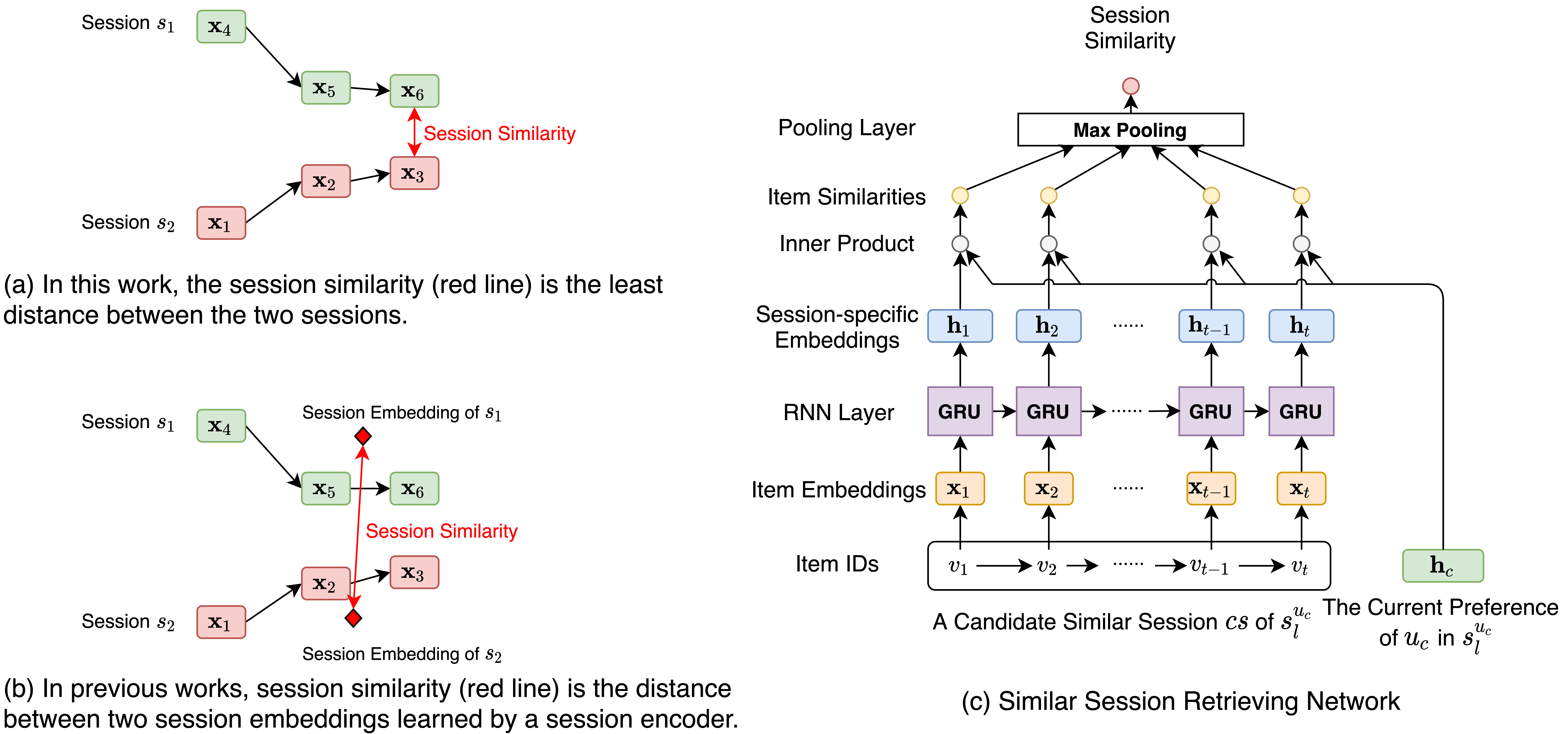}
\caption{(a) and (b) illustrate the difference of session similarity calculation between this work and previous works. The session similarity between $s_1$ and $s_2$ (red line) is the least distance between the two sessions in this work and the distance between the session embeddings generated by a session encoder (e.g., weighted sum of the item embeddings) in previous works. (c) The proposed Similar Sessions Retrieval Network (SSRN).}
\Description{In our work, the session similarity is the maximum similarity between items embeddings in the two sessions. Previous works first use a session encoder to aggregate information of all items in it and represent it with a session embedding, and finally calculate session similarity with the session embeddings. The Similar Session Retrieval Network (SSRN) is based on GRU units and a Max Pooling layer.}
\label{fig_ssrn}
\end{figure}

Introducing collaborative information of other users raises the risk of finding sessions with different user preferences and context, especially when the current session is short and has limited contextual information. 
To find sessions similar to the current session, existing studies \cite{yupu2020neighbor,sun2019can} first encode sessions with fixed-length vectors and then calculate the similarities between sessions based on these vector representations of sessions. 
However, the session similarities obtained may not be accurate because it is difficult for the session encoders to preserve all the information in the sessions. For example, existing works often use the attention-based weighted sum or simply use the mean of item embeddings to represent a session but ignore the positions and orders of the items. 
To address this problem, we \textit{directly measure the similarity of two sessions with the least distance between the embeddings of item from each of the sessions}\footnote{We assume that the embeddings of similar items are ``close'' in the embedding vector space, so the more similar the two items are, the smaller the distance of their embeddings.}. The difference between our idea and previous works is depicted in Figure \ref{fig_ssrn} (a) and (b).
In this way, we can avoid designing a function to encode sessions and the similarity metric between sessions. 


Specifically, as depicted in Figure \ref{fig_ssrn} (c), given a candidate session $cs$ in $\mathbb{H}(u_c)$ or $\mathbb{S}(u_c)$, we first embed each item $v_i\in cs$ to a $d$-dimensional embedding $\mathbf{x}_i$, and then feed them sequentially to an RNN layer. The outputs of the RNN, i.e., $\mathbf{h}_1,\mathbf{h}_2,...,\mathbf{h}_{t}$ contain the information of both the corresponding item and its precedent items in the session. 
Then, for each $\mathbf{h}_{i}$, we calculate its similarity to $\mathbf{h}_c$, i.e., $\lambda_{i,c}$:
\begin{equation}
    \lambda_{i,c}= \mathbf{h}_i \cdot \mathbf{h}_c^T,
\end{equation}
where $i\in [1,t]$ is the position of item in $cs$. 
Finally, the similarity between $cs$ and $C_{ia}$ is obtained by taking the maximum similarity between all the items in $cs$ and $C_{ia}$, which can be seen as the least distance between $\mathbf{h}_c$ and the candidate similar session $cs$:
\begin{equation}
    sim(cs,C_{ia}) = \max_{i\in [1,t]} \lambda_{i,c}.
    \label{eq:sess_sim}
\end{equation}

\subsubsection{Session Encoder}
Once the candidate session $cs$ is similar to $\mathbf{h}_c$, the preference of the user in $cs$ is used to supplement the current session. In this work, we use a attention-based \textit{session encoder} to represent the user preference in $cs$. Specifically, we first embed the user $u_{cs}$ of $cs$ to a $d$-dimensional user preference embedding $\mathbf{\theta}_{cs}$. Then, his preference for an item $v_i$ in $cs$ is calculated by:
\begin{equation}
    \alpha(v_i,u_{cs})=\frac{1}{\eta} \mathbf{x}_i \cdot \mathbf{\theta}_{cs}^T,
\end{equation}
where $\eta=\sum_{j=1}^{t}\mathbf{x}_j\cdot \mathbf{\theta}_{u_{cs}}^T$ is the normalization factor. 
The result user preference of $cs$ is calculated by:
\begin{equation}
    \mathbf{w}_{cs} = \sum_{i=1}^{t} \alpha(v_i,u_{cs}) \times \mathbf{x}_i.
    \label{eq:sess_emb}
\end{equation}


After the session similarity and user preference for each of the sessions in $\mathbb{H}(u_c)$ and $\mathbb{S}(u_c)$ are ready, we aggregate the user preferences of all sessions in the two candidate sets using their similarities to $C_{ia}$ as the weights, respectively and obtain the representations of prior knowledge in Equation \eqref{modulation2}: 
\begin{equation}
\begin{split}
    \beta(\mathbb{H}(u_c)) = MLP_h(\sum_{cs \in \mathbb{H}(u_c)} sim(cs,C_{ia}) \times \mathbf{w}_{cs});\\
    \beta(\mathbb{S}(u_c)) = MLP_s(\sum_{cs \in \mathbb{S}(u_c)} sim(cs,C_{ia}) \times \mathbf{w}_{cs}),
    \label{eq:ctxemb}
\end{split}
\end{equation}
where $MLP_h$ and $MLP_s$ are two Multilayer Perceptron Layers.


\subsection{Prediction Module}

Finally, we use the Prediction Module to modulate the user preference embedding in $C_{ia}$ based on Equation \eqref{modulation2} and the prediction function Equation \eqref{eq:pred2} can be rewritten as:
\begin{equation}
    p(v|C_{ia}, \mathbb{H}(u_c), \mathbb{S}(u_c))=softmax(MLP(\psi(\mathbf{h}_c, \mathbb{H}(u_c), \mathbb{S}(u_c)))).
    \label{eq:pred3}
\end{equation}

\subsection{Optimization and Training}
We train the proposed model with a user-aware mini-batch gradient descent framework based on \cite{ruocco2017inter}. Specifically, for each mini-batch, we select a batch of sessions generated by different users in the dataset. The candidate similar session sets for each user $u_c$ in the batch, i.e., $\mathbb{H}(u_c)$ and $\mathbb{S}(u_c)$ are generated for training.

We treat the prediction as a multi-class classification task and employ the cross-entropy loss to train our model:
\begin{equation}
    \mathcal{L}(v^+)=-[\log p(v^+) + \sum_{v_i\in V, v_i\neq v^+} \log (1-p(v_i))],
\end{equation}
where $v^+$ is the true next item in the current session, and $p(v)$ is short for $p(v|C_{ia}, \mathbb{H}(u_c), \mathbb{S}(u_c))$.

We implement our model using PyTorch\footnote{https://pytorch.org/} and DGL\footnote{https://docs.dgl.ai/} for efficient training and prediction.

\section{Experiments}

\subsection{Data Preparation}

In our experiments, we use two real-world datasets used in previous SBRS works: (1) \textit{Delicious}\footnote{https://grouplens.org/datasets/hetrec-2011/ released in 2nd workshop on information heterogeneity and fusion in recommender systems (HetRec 2011)} \cite{song2019session}, which contains the user tagging actions happened in a social book marking system, and (2) \textit{Reddit}\footnote{https://www.kaggle.com/colemaclean/subreddit-interactions/data/} used in the work \cite{ruocco2017inter}.

We carefully pre-process the datasets by the following approaches in the existing works \cite{ruocco2017inter,ludewig2018evaluation}. First, we remove the users and items with a frequency of less than 10. A user's adjacent interactions are assigned into one session if the time interval (also called inactivity or idle time) between them is less than a threshold, e.g., 3600 seconds \cite{ruocco2017inter}. 
After that, the items not belonging to any sessions and the sessions with a length larger than 20 are removed\footnote{We removed long sessions in order to balance the performance and efficiency of each algorithm. This is a common practice in previous works because the sessions very long in length account for a very small percentage of the datasets,. They may originate from bots or other error reasons. We tested that the performance of some algorithms with longer sessions did not change significantly but increased the running time of these algorithms since they commonly pad short sessions with a fake item to a predefined maximum session length.}. Finally, for each user, we sort his/her sessions by the timestamp, and select the last 10\%, 20\%, 30\% of his/her sessions respectively to form the test set, and the remaining is used as the training set and validation set. The statistics of the datasets are shown in Table \ref{tab:data}. INSERT consistently outperforms the baseline on all proportions so we only show the results w.r.t. 80\%/10\%/10\% training/validation/test splitting in this paper.

\begin{table}
\centering
\caption{Two real-world datasets used in our experiments.}
\begin{tabular}{ccc}
\toprule
      & Delicious & Reddit  \\ 
      \midrule
      \#users        & 1,643      & 18,173   \\
\#items        & 5,005      & 13,521   \\
\#sessions     & 45,603     & 1,119,225 \\
\#interactions & 257,639    & 2,868,050 \\
\#interactions per session & 5.6 & 2.6\\
\#interactions per user & 156.8 & 157.8 \\
\bottomrule
\end{tabular}

\label{tab:data}
\end{table}

\subsection{Experimental Settings}
Following the existing studies \cite{wang2019collaborative,sun2019can}, we use the widely used ranking metrics, i.e., Recall@K and Mean Reciprocal Rank (MRR)@K, to evaluate the recommendation performance in the experiments. 

\subsubsection{Baseline Methods.} We select the following (1) single-session-based (RNN, STAMP, SR-GNN), (2) multi-session-based SBRSs (SKNN, STAN, CSRM, HRNN and II-RNN) and (3) traditional sequential recommender systems (SASRec and BERT4Rec) as baseline methods to compare the recommendation performance of our proposed INSERT model. They are the representative and/or state-of-the-art methods based on a variety of models, such as RNN, attention, memory neural networks and graph neural networks, respectively.

\begin{itemize}
\item \textbf{RNN:} An RNN model built on GRU to extract user preference and sequential patterns in the current session for next-item recommendation in it \cite{ruocco2017inter}.

\item \textbf{STAMP:} An attention and memory neural networks based model aiming to capture users' short-term preferences for next-item recommendations in the current session \cite{liu2018stamp}.

\item \textbf{SR-GNN:} A state-of-the-art single-session-based SBRS built on GNN, in which a gated graph neural network is employed to extract item transition patterns within sessions for next-item recommendations \cite{wu2019session}.

\item \textbf{SKNN:} A KNN-based SBRS which retrieves sessions similar to the current session from the whole dataset to help with the next-item recommendations in the current session \cite{ludewig2018evaluation}.

\item \textbf{STAN:} A multi-session-based SBRS which extends SKNN by considering more information such as the position of items in a session for more accurate next-item recommendations \cite{garg2019sequence}.

\item \textbf{CSRM:}  A multi-session-based SBRS which employs two memory networks to effectively learn the embeddings of both the current session and other sessions for next-item recommendations in the current session \cite{wang2019collaborative}. 

\item \textbf{HRNN:} An SBRS based on a hierarchical RNN which models both a user's historical sessions and her/his current session for next-item recommendations in the current session \cite{quadrana2017personalizing}.

\item \textbf{II-RNN:} An RNN-based SBRS which utilizes the last session of a user's current session to compliment the contextual information of the current session for the next-item recommendations in it \cite{ruocco2017inter}.

\item \textbf{SASRec:} A self-attention based sequential recommender system to model a user's entire sequence \cite{kang2018self}. In this work, we concatenate all the sessions of each user in the training set according to the occurring time to form his training sequence. 

\item \textbf{BERT4Rec:} A sequential recommender system based on bidirectional self-attention networks for next-item prediction. It randomly masks some items and predicts them based on their surrounding items \cite{sun2019bert4rec}. We form the user sequences in the way same as SASRec.

\end{itemize}

\subsubsection{Parameter Settings.} For fair comparisons, we initialize all the hyper-parameters and settings of the baseline methods according to the papers and source codes provided by the authors and then tuned them on the validation set of each dataset for best performance. For each baseline, we test its performance under different values of hyper-parameters including batch size, the number of negative examples, the dimensions of embeddings, and the dimensions of hidden layers, and report the best performance for each baseline method. The number of memory slots in CSRM is set to 256. The number of layers $L=2$ and attention heads $h=2$ in both SASRec and BERT4Rec for best performance. The dimensions of embeddings and hidden states in INSERT are both set to 50. We use the dropout of 20\% to avoid overfitting. An Adam optimizer with an initial learning rate of 0.001 is used to train the INSERT model.

\begin{table*}[ht]
\centering
\caption{Recommendation performance of all compared methods on the short sessions of two real-world datasets. The \textbf{bold numbers} denote the best results, and the second best results are \underline{underlined}.}
\small
\begin{tabular}{c|cccc|cccc}
\toprule
\multicolumn{1}{c|}{} & \multicolumn{4}{c|}{\textbf{Delicious}} & \multicolumn{4}{c}{\textbf{Reddit}}  \\ \hline
   & Recall@5  & Recall@20  & MRR@5    & MRR@20  &  Recall@5  & Recall@20  & MRR@5    & MRR@20  \\ 
   \midrule
RNN                            & 0.1418   & 0.2716    & 0.0830   & 0.0957 & 0.1984  & 0.3544   & 0.1305 & 0.1458 \\
STAMP                          & 0.1476   & 0.2861    & 0.0861  & 0.0997 & 0.1534  & 0.2555   & 0.0981 & 0.1083 \\
SR-GNN                         & 0.1680    & 0.3215    & 0.0931  & 0.1082 & 0.2377  & 0.4016   & 0.1555 & 0.1718 \\ \hline

SKNN                           & 0.1707   & 0.3487    & 0.0780   & 0.0960  & 0.1962  & 0.3758   & 0.0731 & 0.0912 \\
STAN                           & 0.1581   & 0.3150     & 0.0735  & 0.0891 & 0.1894  & 0.3636   & 0.0703 & 0.0879 \\
CSRM                           & 0.1774   & 0.3298    & 0.1007  & 0.1157 & 0.2003  & 0.3609   & 0.1310  & 0.1468 \\

HRNN                           & 0.1749   & 0.3279    & 0.1038  & 0.1189 & 0.3482  & 0.5185   & 0.2436 & 0.2607 \\
II-RNN                          & \underline{0.1846}   & \underline{0.3493}    & \underline{0.1118}  & \underline{0.1279} & 0.3654  & 0.5481   & \underline{0.2533} & \underline{0.2717} 
\\ 
\hline
SASRec & 0.1792 & 0.3431 & 0.0947 & 0.1104 & 0.3219 & \underline{0.5711} & 0.1761 & 0.2012 \\
BERT4Rec & 0.1755 & 0.3143 & 0.1096 & 0.1233 & \textbf{0.4092} & \textbf{0.6231} & 0.2290 & 0.2518 \\
\hline
 INSERT   &              \textbf{0.2163}   &  \textbf{0.3840}     & \textbf{0.1278}  &  \textbf{0.1443} &  \underline{0.3879}  &  0.5588   &  \textbf{0.2684} &  \textbf{0.2858} 
 \\ \bottomrule
\end{tabular}

\label{tab:total}
\end{table*}

\subsection{Recommendation Accuracy Evaluation and Analysis }
We conduct extensive experiments to evaluate our model in terms of accuracy by answering the following questions:
\begin{itemize}
    \item \textbf{Q1:} How does our proposed INSERT model perform when compared with the baseline methods for next-item recommendations in all short sessions?
    \item \textbf{Q2:} How do INSERT and baseline methods perform when making next-item recommendations in short sessions with different lengths?
\end{itemize}

\subsubsection{\textbf{Reply to Q1}: INSERT vs. baselines for next-item recommendations in all short sessions.} To test the performance of all compared methods in short sessions, we ran each experiment 10 times and present the average recommendation accuracy in terms of recall and mean reciprocal rank (MRR) in Table \ref{tab:total}.  

In general, the three single-session-based baseline methods, i.e., RNN, STAMP and SR-GNN, do not perform well due to the limited contextual information used for next-item recommendations in the current session. Next, we specifically analyze the performance of each of these single-session-based baseline methods.  
RNN is easy to make false predictions due to its utilized rigid order assumption over the items within sessions. STAMP performs slightly better than RNN on Delicious but performs the worst on Reddit. This is because the attention mechanism used in STAMP is usually good at handling relatively long sessions, but the majority of sessions in Reddit are very short (less than 4, see the right-side figure in Figure \ref{fig_sesslen}) and thus STAMP cannot perform well. Benefiting from the strong capability in representation learning of GNN, SR-GNN achieves the best performance among the three single-session-based SBRSs, but it only models the item transitions in each single session, and thus its results are worse than the multi-session-based methods.


SKNN and STAN use a simple way to calculate the similarity between sessions, and thus it is hard to find those really similar sessions to improve the next-item recommendations in the current session. Hence, their performance is not so good. With the help of memory networks, CSRM is able to relatively precisely find out those sessions similar to the current session to help with the next-item recommendations in it. Therefore, it performs better than SKNN and STAN. By initializing the current session with the most recent sessions of the same user, HRNN and II-RNN can directly supplement the current session with the personalized context information of the user, leading to performance improvement over CSRM. However, they rely on the assumption that adjacent sessions of the same user need to be highly related, and the models can easily forget the information contained in the sessions far from the current session. Besides, these methods still ignore the sessions from other relevant and similar users to the current user's current session.

SASRec and BERT4Rec supplement the context information of the current session with item sequences of the user formed by concatenating all his sessions according to their occurring time. The recall score of BERT4Rec on Reddit is the best among all the methods, which is probably due to the fact that users on Reddit often visit a few topics over a period of time, so the same item (topic) will often appear repeatedly in the user sequence. 
Except for this, all other metrics of SASRec and BERT4Rec on the two datasets are lower than II-RNN. One reason is that, due to the user cold-start problem in real-world datasets, many users do not have enough historical items and sessions, which may also be short sessions. Another problem of these methods is that concatenating consecutive sessions into one item sequence destroys the intrinsic transaction structure of sessions \cite{wang2019survey}. There are often large time intervals and preference shifts between successive historical sessions of the same user in real-world session-based datasets. Thus, traditional sequential recommender systems may easily make incorrect next-item predictions since they often assume a strict sequential pattern between successive items.




In contrast, INSERT achieves clear improvement w.r.t. recall and MRR on Delicious dataset and Reddit dataset over the best-performing baseline method, respectively (except for Recall of BERT4Rec on Reddit as discussed above). The reason is that, with the carefully devised similar session retrieval module, i.e., SSRN, our proposed INSERT is able to effectively find out those sessions really similar to the current session of the current user from the historical sessions of both the current user and other users. These similar sessions can effectively complement the limited contextual information in the current short session. 


\subsubsection{\textbf{Reply to Q2}: The performance of all methods on short sessions of different lengths.}
\begin{figure*}[t]
\centering
\includegraphics[width=0.8\textwidth]{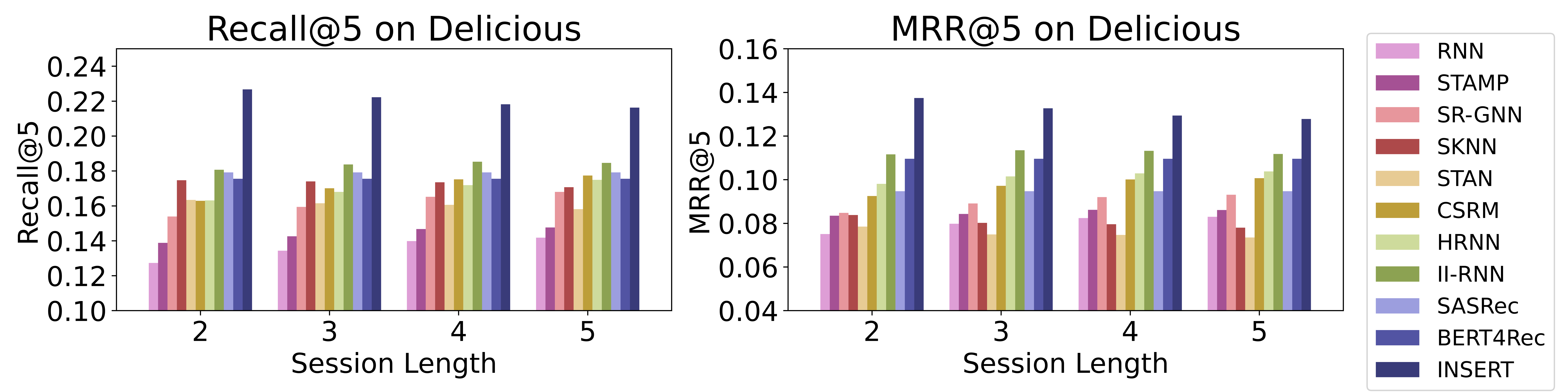} 
\caption{Recommendation performance of all compared methods on short sessions with different lengths on Delicious.}
\Description{These figures show the Recall@5 and MRR@5 on the Delicious dataset of all compared methods on sessions with lengths 2 to 5, respectively. The proposed INSERT performs best in all cases.}
\label{r20slen_d}
\end{figure*}

\begin{figure*}[t]
\centering
\includegraphics[width=0.8\textwidth]{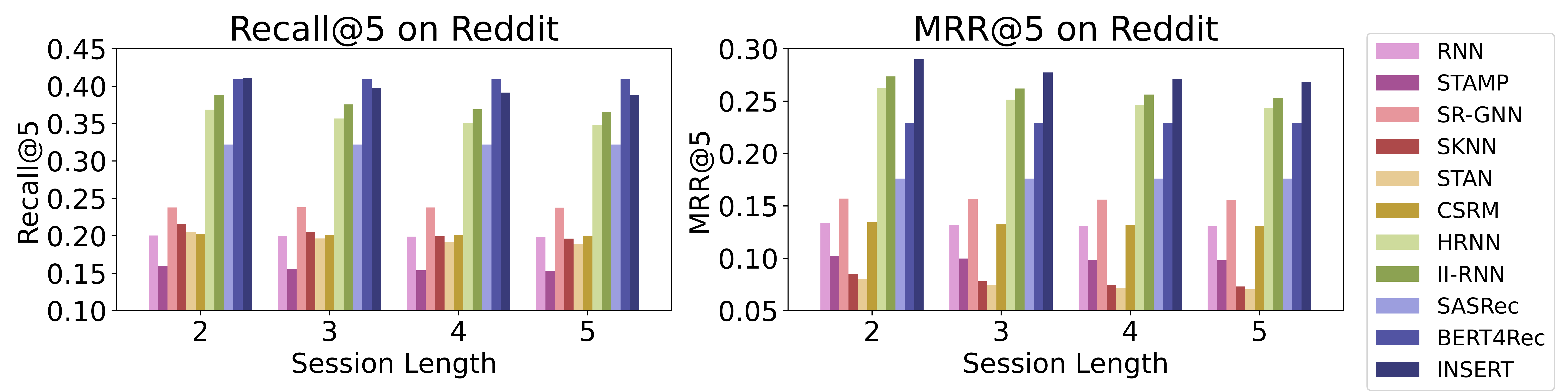} 
\caption{Recommendation performance of all compared methods on short sessions with different lengths on Reddit.}
\Description{These figures show the Recall@5 and MRR@5 on the Reddit dataset of all compared methods on sessions with lengths 2 to 5, respectively. The proposed INSERT performs best in terms of MRR@5 and has the second-best Recall@5.}
\label{r20slen_r}
\end{figure*}

Figure \ref{r20slen_d} and \ref{r20slen_r} present the recall@5 and MRR@5 of all compared methods on Delicious dataset and Reddit dataset, respectively. Two observations can be drawn from them: (1) our proposed INSERT performs the best on short sessions with different lengths in most cases, and (2) the shorter the sessions, the more significant the performance improvement of INSERT over the baseline models. Therefore, INSERT is better at improving the performance of next-item prediction in short sessions. Both observations clearly demonstrate the stronger capability of INSERT in making next-item recommendations in short sessions compared with the baseline methods.



\subsection{Ablation Analysis}
To verify the effectiveness of different modules we designed in INSERT, we conduct experiments to answer the following two questions for ablation analysis.   
\begin{itemize}
    \item \textbf{Q3:} Can the INSERT modules benefit the next-item recommendations in short sessions? 
    \item \textbf{Q4:} How does the SSRN module performs when compared with other session similarity calculation methods?
\end{itemize}

\subsubsection{\textbf{Reply to Q3}: the performance of three simplified versions of INSERT} We implement three simplified versions of INSERT: (1) \textbf{INSERT-c}, which is simplified from INSERT by removing the global module and thus makes next-item recommendations based on a user's current session only; (2) \textbf{INSERT-h}, which is simplified from INSERT by removing the prior knowledge from other users, i.e., $\mathbb{S}$, while keeping $\mathbb{H}$ to retrieve similar sessions from the historical sessions of the current user to help with the next-item recommendations in the current session of the current user; and (3) \textbf{INSERT-o} which is simplified from INSERT by removing the modulation based on the sessions of the current user, i.e., $\mathbb{H}$, while keeping $\mathbb{S}$ to retrieve similar sessions from the historical sessions of other users similar to the current user to complement the contextual information.  

The experimental results of the three simplified versions and the full version, namely INSERT, are shown in Table \ref{tab:ablation}. 
From Table \ref{tab:ablation}, we can see that INSERT-c does not perform well because it only utilizes the limited contextual information in the current short session for next-item recommendations. INSERT-h performs better than INSERT-c because it obtains sessions similar to the current session from the historical sessions of the current user to complement the limited contextual information in the current session. INSERT-o achieves slightly better results than INSERT-h, showing that the historical sessions from other users similar to the current user can play an important role in next-item recommendations in short sessions. 
INSERT performs the best because it finds out those sessions similar to the current session of the current user from historical sessions of both the current user and other users.



\begin{table}[ht]
\centering
\caption{Recommendation performance of INSERT and its simplified versions on Delicious.} 
\small
\begin{tabular}{lcccc}
\toprule
 & Recall@5 & Recall@20 & MRR@5  & MRR@20 \\ \midrule
INSERT-c  & 0.1418   & 0.2716    & 0.0830  & 0.0957 \\
INSERT-h  & 0.1833   & 0.3407    & 0.1101 & 0.1254 \\
INSERT-o & 0.1975   & 0.3629    & 0.1160 & 0.1322     \\
INSERT-a & 0.1891   & 0.3508    & 0.1102 & 0.1259     \\
INSERT     & \textbf{0.2163}   &  \textbf{0.3840}     & \textbf{0.1278}  &  \textbf{0.1443} \\ \bottomrule
\end{tabular}

\label{tab:ablation}
\end{table}

\subsubsection{\textbf{Reply to Q4}: performance of SSRN and other alternative session similarity calculation methods.} We implement a variant of INSERT called \textbf{INSERT-a} by replacing the SSRN module in INSERT with the similarity calculation method used in \cite{yupu2020neighbor}. INSERT-a simply represents a given session with the mean value vector of the embeddings of items in the session and then calculates the similarity between sessions using inner product operation between their representations. The recommendation performance of INSERT-a is shown in Table \ref{tab:ablation} as well. It is clear that INSERT equipped with SSRN achieves better performance than INSERT-a. This shows that given a current short session, SSRN can effectively calculate more precise similarity values between it and other sessions. This contributes to precisely find out those sessions in the dataset that are similar to the current session.

\section{Conclusions}
Targeting the challenging problem of next-item recommendation in short sessions with limited items, in this paper, we first formulate the problem as a few short learning (FSL) problem and then devise a novel inter-session collaborative recommender network (INSERT) inspired by the basic idea of a representative approach for FSL, i.e., meta learning. With carefully designed local module and global module, INSERT can not only capture a user's timely preference from the current short session but also modulate and optimize such preference with complementary prior knowledge learned from other similar sessions that are precisely retrieved from the whole dataset. As a result, INSERT is able to accurately recommend the next item in the current short session based the optimized user's preference. Extensive experiments conducted on two real-world datasets show the superiority of INSERT over state-of-the-art SBRSs in predicting next-item in short sessions. In the future, we will explore how to devise more powerful global module to more effectively extract useful prior knowledge from the whole dataset to better complement the very limited preference information embedded in the current short session for more accurate next-item recommendations.  

\begin{acks}
This work was funded by China Scholarship Council under grant No. 201906170208, and was partially supported by Australian Research Council Discovery Project DP200101441, the National Key Research and Development Program of China No. 2020YFA0714103, the Innovation Capacity Construction Project of Jilin Province Development and Reform Commission No. 2019C053-3, and the Science \& Technology Development Project of Jilin Province No. 20190302117GX.
\end{acks}

\bibliographystyle{ACM-Reference-Format}
\bibliography{main}

\end{document}